\documentclass[showpacs,preprintnumbers,amsmath,amssymb]{revtex4}

\usepackage{epsfig}% Include figure files
\usepackage{dcolumn}% Align table columns on decimal point
\usepackage{bm}% bold math
\usepackage{citesort}
\date{\today}
\newcommand{\bmat}{\left(\begin{array}}

\newcommand{\emat}{\end{array}\right)}
\newcommand{\be}{\begin{equation}}
\newcommand{\ee}{\end{equation}}
\newcommand{\bea}{\begin{eqnarray}}
\newcommand{\eea}{\end{eqnarray}}

\def\lsim{\raise0.3ex\hbox{$\;<$\kern-0.75em\raise-1.1ex\hbox{$\sim\;$}}}
\def\gsim{\raise0.3ex\hbox{$\;>$\kern-0.75em\raise-1.1ex\hbox{$\sim\;$}}}

\def\OMIT#1{}

\newcommand{\nn}{\nonumber}

\newcommand{\plus}{\ensuremath{\! + \!}}
\newcommand{\minus}{\ensuremath{\! - \!}}

\begin{document}
\renewcommand{\thefootnote}{\fnsymbol{footnote}}

\baselineskip 16pt
\title{Supersymmetric contribution to $ B\to \rho K$ and  $
B\to \pi K^*$ decays in $SCET$}
\author{Gaber Faisel}
\affiliation{ Department of Physics and Center for Mathematics and Theoretical Physics,
National Central University, Chung-li, TAIWAN 32054.}

\affiliation{Egyptian Center for Theoretical Physics, Modern
University for Information and Technology, Cairo, Egypt}
\author{David Delepine}
\affiliation{Departamento de F\'isica, DCI, Campus Le\'on, Universidad
  de Guanajuato, C.P. 37150, Le\'on, Guanajuato, M\'exico}
\author{M. Shalaby}
\affiliation{Ain Shams University, Faculty of
Science, Cairo 11566, Egypt.}

\begin{abstract}

We analyze the supersymmetric contributions to the direct CP
asymmetries  of the decays $B \to \pi K^*$ and $B\to\rho K$ within
Soft Collinear Effective Theory. We extend the Standard Model
analysis of these asymmetries to include the next leading order
QCD corrections. We find that, even with QCD correction, the
Standard Model predictions can not accommodate the direct CP
asymmetries in these decay modes. Using Mass Insertion
Approximation (MIA), we show that non-minimal flavor SUSY
contributions mediated by gluino exchange can enhance the CP
asymmetries significantly and thus can accommodate the
experimental results.
\end{abstract}
\maketitle
%%%%%%%%%%%%%%%%%%%%%%%%%%%%%%%%%%
\section{ Introduction}

 In the standard model (SM), Charge conjugation Parity (CP) violation and flavour
transition arise from the complex Yukawa couplings in the Cabibbo
Kobayashi Maskawa (CKM) matrix. The effect of this phase has been
first observed in kaon system and confirmed in B decays. However,
the expected CP asymmetries in some decay channels for B meson are
in contradiction with the experimental measurements carried by
Babar and Belle B-factories and proton antiproton collider as
Tevatron, with its experiments CDF and D0. The largest discrepancy
has been observed in the decay $B \to K \pi$ where the world
averages for the CP asymmetries of $B^0 \to K^{\pm} \pi^{\mp}$ and
$B^{\pm} \to K^{\pm} \pi^0$ are  given
by\cite{TheHeavyFlavorAveragingGroup:2010qj}:\bea
{\cal A}_{CP}(B^0 \to K^{\pm} \pi^{\mp}) &=& -0.098\pm 0.012 ,\\
{\cal A}_{CP}(B^{\pm} \to K^{\pm} \pi^0) &=& 0.050\pm 0.025 .%
\eea which implies that \begin{equation} \Delta {\cal
A}_\mathrm{CP}={\cal A}_{CP}(B^{\pm} \to K^{\pm} \pi^0) -{\cal
A}_{CP}(B^0 \to K^{\pm} \pi^{\mp})=0.14 \pm 0.029,
\end{equation}   In the SM and using QCD factorization approach, the results
of the above two asymmetries  read \cite{Beneke:2003zv}: \bea
{\cal A}_{CP}(B^{\pm} \to K^{\pm} \pi^0)& = & \left(
7.1^{+1.7+2.0+0.8+9.0}
_{-1.8-2.0-0.6-9.7} \right) \% \\
{\cal A}_{CP}(B^0 \to K^{\pm} \pi^{\mp}) & = & \left(
4.5^{+1.1+2.2+0.5+8.7}_ {-1.1-2.5-0.6-9.5} \right) \% \; , \eea
where the first error corresponds to uncertainties on the CKM
parameters and the other three errors correspond to variation of
various hadronic parameters. These results imply that $\Delta
{\cal A}^{QCD}_\mathrm{CP}= 0.025 \pm 0.015$, which differs from
the experimental value by 3.5$\sigma$ and thus motivate exploring new physics beyond SM.

  The decay modes $B \to \pi K^*$ and $B\to\rho K$ are generated at
the quark level in  the same way as $B \to K\pi$ and  hence it is
interesting  to explore hints of New Physics (NP) in these decays.
These decay modes are studied within  SM in framework of
QCDF~\cite{Beneke:2003zv},
PQCD~\cite{Lu:2000hj,Liu:2005mm,Guo:2006uq,Guo:2007vw} and Soft
Collinear Effective Theory (SCET) \cite{Wang:2008rk}. A detailed
comparison between the results for the branching ratios and CP
asymmetries in these different factorizations methods can be found
in Ref.(\cite{Wang:2008rk}). The comparison showed that PQCD
results for most $B \to \pi K^*$ and $B\to\rho K$ channels are
much larger than SCET results. On the other hand the QCDF results
are small and comparable with SCET results but with a relative
minus sign. Moreover, in SCET, the direct CP asymmetries of $B^-
\to \pi^- \bar{K}^{*\,0}$ and $B^-\to\rho^- \bar{K}^0$ are zero
while the CP asymmetries in other channels are small. Recently, in
Ref.(\cite{Chiang:2009hd}) fits to  $B \to \pi K^*$ and $B\to\rho
K $ decays are performed where data can be accommodated within the
standard model due principally to the large experimental
uncertainties, particularly in the CP-violating asymmetries.

One of the four large experiments operating at the Large Hadron
Collider (LHC) is LHCb. The main task of the LHCb is to measure
precisely the CP asymmetries in B meson decays.  These
measurements are so important to test the different mechanisms
proposed by  many models beyond SM to explain the
matter-antimatter asymmetry.  This test can be regarded as an
indirect search for physics beyond SM.

  Supersymmetry (SUSY) is one of the most interesting candidates for physics
beyond the standard model as it naturally solves the hierarchy
problem. In addition, SUSY has new sources for CP violation which
can account for the baryon number asymmetry and affect other CP
violating observables in the B and K decays. The effects of these
phases on the CP asymmetries in semi-leptonic $\tau$ decays has
been studied in
Refs.(\cite{Delepine:2006fv,Delepine:2007qg,Delepine:2008zzb}).

In this paper, we analyze the SUSY contributions to the CP
asymmetries of the $B \to \pi K^*$ and $B\to\rho K$ decays in the
framework of SCET\cite{Bauer:2000ew,Bauer:2000yr,Chay:2003zp,Chay:2003ju}. SCET  is an
effective field theory describing the dynamics of highly energetic particles moving close to the
light-cone interacting with a background field of soft quanta\cite{Fleming:2009fe}.
It provides a systematic and  rigorous way to deal with the
decays of the heavy hadrons that involve different energy scales. The
scaling of fields and momenta in SCET depends on a small parameter
$\lambda $. Generally $\lambda $ is defined as the ratio of the smallest  and
the largest energy scales in the given process. Then, the SCET Lagrangian and effective
Hamiltonian are expanded in terms of $\lambda$ that help to reduce the complexity of the calculations.
In addition, the factorization formula provided by SCET is
perturbative to all powers in $\alpha_s$ expansion.

This paper is organized as follows. In Sec.~\ref{sec:formalism},
we briefly review the decay amplitude for $B \to M_1M_2$ within
SCET framework. Accordingly, we analyze the CP asymmetries and
branching ratios for $B \to \pi K^*$ and $B\to\rho K$ within SM in
Sec.~\ref{sec:SM}. In Sec.~\ref{sec:SUSY}, we discuss the SUSY
contributions  to the CP asymmetries of the $B \to \pi K^*$ and
$B\to\rho K$ decays.  We give our conclusion in
Sec.~\ref{sec:conclusion}.

\section{ $ B\to {M_1M_2}$ in $ SCET $ }\label{sec:formalism}

The amplitude of $ B\to {M_1M_2}$ where $M_1$ and $M_2$ are light
mesons  in SCET can be written as follows \bea{\cal A}_{B\to
M_1M_2}^{SCET}
 &=& {\cal A}_ {B\to M_1M_2}^{LO} +  {\cal A}_{B\to M_1M_2}^ {\chi} +
 {\cal A}_{B\to M_1M_2}^{ann} +  {\cal A}_{B\to M_1M_2}^{c.c} \eea

Here ${\cal A}_ {B\to M_1M_2}^{LO}$ denotes the leading order
amplitude in the expansion $1/m_b$, $ {\cal A}_{B\to M_1M_2}^
{\chi}$ denotes the chirally enhanced penguin amplitude,  $ {\cal
A}_{B\to M_1M_2}^{ann}$ denotes the annihilation amplitude and $
{\cal A}_{B\to M_1M_2}^{c.c}$ denotes the long distance  charm
penguin contributions. In the following we give a brief account
for each amplitude.

\subsection{Leading order amplitude}

At leading power in $(1/m_b)$ expansion, the full QCD effective
weak Hamiltonian of the $\Delta_B=1$ decays is matched into the
corresponding weak Hamiltonian in $SCET_I$ by integrating out  the
hard scale $ m_b $. Then, the $SCET_I$ weak Hamiltonian is
matched into the weak Hamiltonian $SCET_{II}$ by integrating out
the hard collinear modes with $p^2\sim \Lambda m_b$ and the
amplitude of the $\Delta_B=1$ decays at leading order in $\alpha_s$ expansion
can be obtained via~\cite{Bauer:2002aj}:

 \begin{eqnarray}
 {\cal A}^{LO}_{B\rightarrow M_1M_2}&=& - i\big\langle
M_1M_2 \big| H^{SCET_{II}}_W \big|
\bar{B}\big\rangle\nonumber\\&=&\frac{G_F
m_B^2}{\sqrt{2}}\Big(f_{M_1}[\int^1_0
 dudzT_{M_1J}(u,z)\zeta^{B M_2}_J(z)\phi_{M_1}(u)\nonumber\\
  &+& \zeta^{B M_2}\int^1_0
 du T_{M_1\zeta}(u)\phi_{M_1}(u)]+(M_1 \leftrightarrow
 M_2)\Big).\label{amp1}
 \end{eqnarray}

At leading order in $\alpha_s$ expansion, the parameters $\zeta^{B
(M_1,M_2)}$, $\zeta_J^{B (M_1,M_2)}$ are treated
 as hadronic parameters and can be determined through the $\chi^2$ fit method using  the
 non leptonic decay  experimental data of  the branching fractions and CP
 asymmetries. At first order in $\alpha_s$ expansion, $\zeta^{BM}_J(z)$ can be written as a polynomial in z
as follows \cite{Jain:2007dy}
  \be \label{zJ1}
\zeta_J^{B M}(z) =  2z\, \zeta_J^{B M}  - A_{1}^{B M}(4z -6 z^2 )
  + \frac{5}{6} A_2^{B M} (z - 6 z^2 + 6 z^3). \,
\ee
where again $\zeta_J^{B M}$ are treated as hadronic parameters
which are determined through the fit to the non leptonic decay
 data.  The hard kernels $T_{(M_1,M_2)\zeta}$ and $T_{(M_1,M_2) J}$ are expressed in terms of
 $c_i^{(f)}$ and $b_i^{(f)}$ which are functions of the
 Wilson coefficients  as follows \cite{Jain:2007dy}
 \begin{eqnarray} \label{Tz}
 T_{1\zeta}(u) &=
  {\cal C}_{u_L}^{BM_2} \, {\cal C}_{f_Lu}^{M_1} \, c_1^{(f)}(u)
  +   {\cal C}_{f_L}^{BM_2} \, {\cal C}_{u_Lu}^{M_1} \, c_2^{(f)}(u)
  \nn \\
 & +   {\cal C}_{f_L}^{BM_2} \, {\cal C}_{u_Ru}^{M_1} \, c_3^{(f)}(u)
  +  {\cal C}_{q_L}^{BM_2} \, {\cal C}_{f_Lq}^{M_1} \, c_4^{(f)}(u)
  ,\nn\\
 T_{1 J}(u,z) &=
   {\cal C}_{u_L}^{BM_2} \, {\cal C}_{f_Lu}^{M_1} \, b_1^{(f)}(u,z)
  +   {\cal C}_{f_L}^{BM_2} \, {\cal C}_{u_Lu}^{M_1} \, b_2^{(f)}(u,z)
   \nn\\
 &+   {\cal C}_{f_L}^{BM_2} \, {\cal C}_{u_Ru}^{M_1} \, b_3^{(f)}(u,z)
  +  {\cal C}_{q_L}^{BM_2} \, {\cal C}_{f_Lq}^{M_1} \, b_4^{(f)}(u,z).
\end {eqnarray}
here $f$ stands for $d$ or $s$ and  ${\cal C}_i^{BM}$ and ${\cal
C}_{i}^M$ are Clebsch-Gordan coefficients that depend on the
flavor content of the final states. For instance, we have ${\cal
C}_{u_L}^{\bar B^0\pi^+}=+1$, ${\cal C}_{d_Lu}^{\pi^-}=+1$, ${\cal
C}_{d_Ru}^{\pi^-}=-1$, ${\cal C}_{u_L}^{\bar B^0\rho^+}=+1$, and
${\cal C}_{d_L u}^{\rho^-}= {\cal C}_{d_R u}^{\rho^-}=+1$, ${\cal
C}_{d_L}^{B^-\pi^-}=+1$ and ${\cal C}_{u_ru}^{\pi^0}=-
\frac{1}{\sqrt{2}}$ and $c_i^{(f)}$ and $b_i^{(f)}$   are given by
\cite{Bauer:2004tj}
\begin{eqnarray}\label{cmatch}
c_{1,2}^{(f)}&=&\lambda_u^{(f)}\Big[C_{1,2}+\frac{1}{N}C_{2,1}\Big]
-\lambda_t^{(f)}\frac{3}{2}\Big[\frac{1}{N}C_{9,10}
+C_{10,9}\Big]+ \Delta c_{1,2}^{(f)},\nonumber\\
c_3^{(f)}&=&-\frac{3}{2}\lambda_t^{(f)}\Big[C_7+\frac{1}{N}C_8\Big]
+ \Delta c_3^{(f)},\nonumber\\
c_4{(f)}&=&-\lambda_t^{(f)}\Big[\frac{1}{N}C_3+C_4
-\frac{1}{2N}C_9-\frac{1}{2}C_10\Big]+ \Delta c_4^{(f)},
\end{eqnarray}
and
\begin{eqnarray}
b_{1,2}^{(f)}&=&\lambda_u^{(f)}\Big[C_{1,2}+\frac{1}{N}
\Big(1-\frac{m_b}{\omega_3}\Big)C_{2,1}\Big]-
\lambda_t^{(f)}\frac{3}{2}\Big[C_{10,9}+\frac{1}{N}
\Big(1-\frac{m_b}{\omega_3}\Big)C_{9,10}\Big]+ \Delta b_{1,2}^{(f)},\nonumber\\
b_3^{(f)}&=&-\lambda_t^{(f)}\frac{3}{2}\Big[C_7
+\Big(1-\frac{m_b}{\omega_2}\Big)\frac{1}{N}C_8\Big]+ \Delta b_3^{(f)},\nonumber\\
b_4^{(f)}&=&-\lambda_t^{(f)}\Big[C_4+\frac{1}{N}
\Big(1-\frac{m_b}{\omega_3}\Big)C_3\Big]+\lambda_t^{(f)}\frac{1}{2}
\Big[C_{10}+\frac{1}{N}\Big(1-\frac{m_b}{\omega_3}\Big)C_9\Big]+
\Delta b_4^{(f)},
\end{eqnarray}
where $\omega_2=m_bu$ and $\omega_3=-m_b\bar{u}$. $u$ and
$\bar{u}=1-u $ are momentum fractions for the quark and antiquark
$\bar{n}$ collinear fields. The $\Delta c_i^{(f)}$ and $\Delta
b_i^{(f)}$ denote terms depending on $\alpha_s$ generated by
matching from $H_W$. The ${\cal O}(\alpha_s)$ contribution to
$\Delta c_i^{(f)}$ has been calculated
in Refs.(\cite{Beneke:1999br,Beneke:2000ry,Chay:2003ju}) and later in Ref.
(\cite{Jain:2007dy}) while the ${\cal O}(\alpha_s)$contribution to
$\Delta b_i^{(f)}$ has been calculated
in Refs.(\cite{Beneke:2005vv,Beneke:2006mk,Jain:2007dy}).

\subsection{ Chirally enhanced penguins amplitude }

 Corrections of order $\alpha_s(\mu_h)(\mu_M\Lambda/m_b^2)$ where $\mu_M$ is the chiral
scale parameter generate the so called Chirally enhanced penguins
amplitude $ {\cal A}_{B\to M_1M_2}^ {\chi}$\cite{Jain:2007dy}.
$\mu_M$ for kaons and pions can be of order $(2GeV)$ and therefore
chirally enhanced terms can compete with the order
$\alpha_s(\mu_h)(\Lambda/m_b)$ terms.   The chirally enhanced
amplitude for  $B\to M_1M_2$ decays is given by\cite{Jain:2007dy}

\begin{eqnarray}\label{chienhanced} A^\chi(\bar B\to M_1 M_2) &=& \frac{G_F
m_B^2}{\sqrt2}  \bigg\{- \frac{\mu_{M_1}  f_{M_1}}{3m_B}
\zeta^{BM_2}  \int_0^1 du R_{1}(u)\phi_{pp}^{M_1}(u)
   + (1\leftrightarrow 2)\nonumber\\
   &-&  \frac{\mu_{M_1} f_{M_1}}{3m_B} \int_0^1 du dz
   R_{1}^{J}(u,z) \zeta_J^{BM_2}(z) \phi_{pp}^{M_1}(u)
 + (1\leftrightarrow 2) \nonumber\\ & -& \frac{\mu_{M_2}f_{M_1}}{6 m_B}\int_0^1 dudz
   R_{1}^{\chi}(u,z) \zeta_{\chi}^{BM_2}(z) \phi^{M_1}(u)
 +(1\leftrightarrow 2) \bigg\}
 \end{eqnarray}

The factors $\mu_M$ are generated by pseudoscalars and so they
vanish for vector mesons~\cite{Jain:2007dy}. The pseudoscalar
light cone amplitude $\phi_{pp}^M(u)$ is defined as
\cite{Hardmeier:2003ig,Arnesen:2006vb}
\begin{equation}
\phi_{pp}^P(u) =3u[\phi_p^P(u)+\phi_\sigma^{P\prime}(u)/6+ 2
f_{3P}/(f_P \mu_P) \int dy'/y' \phi_{3P}(y-y',y)].
 \end{equation}

 $\phi_{pp}^M$  are commonly expressed in terms of the first few terms in the Gegenbauer series
 \be
\phi_{pp}^M(x) =  6 x(\!1\!-\!x\!) \big\{1+
a_{1pp}^{M}(\!6x\!-\!3)+ 6 a_{2pp}^{M}(\!1\!-\!5 x\! +\!5 x^{2})
\big\} \,. \ee

As before, following the same procedure for treating
$\zeta^{BM}_J(z)$ we take  $\zeta^{BM}_{\chi}(z)$ as
\cite{Jain:2007dy} \be \zeta_{\chi}^{BM}(z) =
  2z \zeta_{\chi}^{BM} - A_{\chi 1}^{BM}(4z -6 z^2 )
  + \frac{5}{6} A_{\chi 2}^{BM} (z - 6 z^2 + 6 z^3).
\ee

 The hard kernels $R_{K},R_{\pi},R_{K}^J,R_{\pi}^J,R_{K}^{\chi}$
and $ R_{\pi}^{\chi}$ can be expressed in terms of Clebsch-Gordan
coefficients for the different final states as\cite{Jain:2007dy}
\begin{eqnarray}
\label{chihard} R_{1}(u) &=  {\cal C}_{q_R}^{BM_{2}} {\cal
C}_{f_Lq}^{M_{1}}
  \Big[ c^{\chi}_{1(qfq)}+\frac{3}{2}e_q \,c^{\chi}_{2(qfq)} \Big]
\,, \\
R_{1}^J(u,z) &=  {\cal C}_{q_R}^{BM_{2}} {\cal C}_{f_Lq}^{M_{1}}
  \Big[ b^{\chi}_{3(qfq)}+\frac{3}{2}e_q \,b^{\chi}_{4(qfq)} \Big]
\,, \nn \\
R_{1}^{\chi}(u,z) &=
  {\cal C}_{q_L}^{BM_{2}} {\cal C}_{f_Lq}^{M_{1}} \, b^{\chi}_{1(qfq)}
 + {\cal C}_{u_L}^{BM_{2}} {\cal C}_{f_Lu}^{M_{1}}\,  b^{\chi}_{1(ufu)} \nn \\
& +{\cal C}_{f_L}^{BM_{2}} {\cal C}_{u_L u}^{M_{1}}\,
b^{\chi}_{1(fuu)}
  +{\cal C}_{f_L}^{BM_{2}} {\cal C}_{u_R u}^{M_{1}}\, b^{\chi}_{2(fuu)} \, .\nn
\end{eqnarray}
Summation over $q=u,d,s$ is implicit and  $c_{i}^{\chi}$ and
$b_{i}^{\chi}$ are expressed in terms of the short-distance
Wilson coefficients  as \cite{Jain:2007dy}
\begin{eqnarray}
 c_{1(qfq)}^{\chi}&=& \lambda_t^{(f)}
\Big(C_6 \plus \frac{C_5}{N_c} \Big) \frac{1}{u\bar u} + \Delta
c_{1(qfq)}^{\chi}\nonumber\\
c_{2(qfq)}^{\chi} &=&\lambda_t^{(f)} \Big(C_8 \plus
\frac{C_7}{N_c} \Big) \frac{1}{u\bar u} + \Delta c_{2(qfq)}^{\chi}\nonumber\\
 b_{1(qfq)}^{\chi} &=& 2
\lambda_t^{(f)} \bigg[\frac{(1 \plus u z)}{u z}
 \Big(\frac{C_3}{N_c} \minus \frac{C_9}{2N_c}\Big)\plus
  C_4 \minus \frac{C_{10}}{2}\bigg]+ \Delta  b_{1(qfq)}^{\chi}\nonumber\\
b_{2(fuu)}^{\chi} &=& 3 \lambda_t^{(f)} \bigg[ C_7\plus
 \frac{C_8}{N_c}  -  \frac{1}{\bar u z}\frac{C_8}{N_c}
  \bigg]  +  \Delta b_{2(fuu)}^{\chi}\nonumber\\
b_{3(qfq)}^{\chi} &=& \lambda_t^{(f)}  \frac{1}{u\bar u }
   \Big(C_6\plus \frac{C_5}{N_c} \Big) +\Delta b_{3(qfq)}^{\chi}\nonumber\\
b_{4(qfq)}^{\chi} &=& \lambda_t^{(f)} \frac{1}{u\bar u }
    \Big( C_8 \plus \frac{C_7}{N_c} \Big)+ \Delta b_{4(qfq)}^{\chi}\nonumber\\
b_{1(ufu)}^{\chi} &=& \frac{2(1\plus uz)}{u z}\Big(\minus
\frac{C_2}{N_c} \lambda_u^{(f)}
    \plus \frac{3C_9}{2N_c} \lambda_t^{(f)}  \Big)\minus \Big( 2 C_1 \lambda_u^{(f)}
  \minus  3 C_{10} \lambda_t^{(f)} \Big)+ \Delta  b_{1(ufu)}^{\chi}\nonumber\\
 b_{1(fuu)}^{\chi} &=& \frac{2(1\plus uz)}{u z}
    \Big(\minus\lambda_u^{(f)}  \frac{C_1}{N_c}
    \plus  \lambda_t^{(f)} \frac{3C_{10}}{2N_c}\Big) - \Big(2C_2 \lambda_u^{(f)}
   \minus 3C_{9} \lambda_t^{(f)} \Big)+\Delta
   b_{1(fuu)}^{\chi}\nonumber\\
   \end{eqnarray}

   The $\Delta c_{i}^{\chi}$ and $\Delta b_{i}^{\chi}$ terms denote perturbative
corrections that  can be found in Ref.(\cite{Jain:2007dy}).

\subsection{ Annihilation amplitudes}

 Annihilation amplitudes ${\cal A}_{B\to M_1M_2}^{ann}$ have been studied
 in PQCD and QCD factorization in Refs.(\cite{Keum:2000ph,Lu:2000em,Beneke:2001ev,Kagan:2004uw}).
Within SCET, the annihilation contribution becomes factorizable
and  real at leading order,$ {\cal
O}(\alpha_s(m_b)\Lambda/m_b)$\cite{Manohar:2006nz}. In our
numerical calculation, we do not include the contributions from
penguin annihilation  as their size is small and contains  large
uncertainty compared to the other
contributions\cite{Arnesen:2006vb,Jain:2007dy}.

\subsection{ Long distance  charm penguin amplitude }

  The long distance  charm penguin amplitude $ {\cal A}_{B\to M_1 M_2}^{c.c}$ is given as
follows \be {\cal A}_{B\to M_1 M_2}^{c.c}=|{\cal A}_{B\to M_1
M_2}^{c.c}|e^{i \delta_{cc}} \ee where $\delta_{cc}$ is the strong
phase of the charm penguin. The modulus and the phase of the charm penguin
are fixed through the fitting with non leptonic decays in a
similar way to the hadronic parameters $\zeta^{B (M_1,M_2)}$,
$\zeta_J^{B (M_1,M_2)}$.

\section{SM contribution to the CP asymmetries and branching ratios  of $B\to \pi K^*$ and $B\to
\rho K$ decays}\label{sec:SM}
In this section, we analyze the SM contribution to the CP
asymmetries and the branching ratios for $B\to \pi K^*$ and $B\to
\rho K$ decays. We follow ref.~\cite{Jain:2007dy} and work in the
next leading order of $\alpha_s$ expansion. We take
$\alpha_s(m_Z)=0.118$, $m_t=170.9\,{\rm GeV}$, $m_b=4.7\,{\rm
GeV}$  and the Wilson coefficients $C_i$  can be found in
Ref.(\cite{BBL}). For the other hadronic parameters, we use the same input values given
in Ref.(\cite{Jain:2007dy}). For the charm penguin parameters we use
the values listed in Ref.(\cite{Wang:2008rk}).

\begin{table}
 \begin{center}
\begin{tabular}{|c|c|c|c|c|c|}
  \hline
    Decay channel&  Exp.  &  SM prediction   \\
  \hline
 $\pi^0 K^{(*)\,+}$  & 6.9 $\pm 2.3$ & $ 7.2_{-0.2-0.9}^{+0.3+1.1} $     \\\hline
    $\pi^- K^{(*)\,+}$ & $ 8.6\pm 0.9$& $  7.8_{-0.2-1.0}^{+0.2+1.1}$      \\\hline
  $ \pi^0 \bar{K}^{(*)\,0}$ & $2.4\pm 0.7$& $  7.8_{-0.5-1.0}^{+0.5+1.2} $    \\\hline
  $ \pi^+ \bar{K}^{(*)\,0}$ & $9.9^{+ 0.8}_{-0.9}$& $ 10.3_{-0.7-1.4}^{+0.7+1.7}$     \\\hline
$\rho^0 K^{+}$&$3.81^{+0.48}_{-0.46}$ & $  4.8_{-0.6-0.7}^{+0.6+0.8}$  \\\hline
$ \rho^+ \bar{K}^{0}$&$8.0^{+1.5}_{-1.4}$ &  $  10.9_{-0.6-1.5}^{+0.6+1.7}$   \\\hline
$ \rho^0 \bar{K}^{0}$& $4.7\pm 0.7$ & $  10.2_{-0.6-1.4}^{+0.6+1.6} $   \\\hline
$\rho^- K^{+} $& $8.6^{+0.9}_{-1.1}$& $ 2.6_{-0.4-0.4}^{+0.5+0.4}$  \\
\hline
\end{tabular}
 \end{center}
\caption{Branching ratios in units $10^{-6}$ of $B\to \pi K^*$ and
$B\to \rho K$ decays. For comparison, we list the experimental
results given in Ref.\cite{TheHeavyFlavorAveragingGroup:2010qj}. The
first uncertainty in the predictions is due to the  uncertainties in SCET parameters
while the second uncertainty  is due to  the  uncertainties in the CKM
matrix elements. }\label{branch}
\end{table}

The decay modes $B\to \pi K^*$ and $B\to \rho K$ are generated at the quark level
via $b\rightarrow s $  transition  and thus we can decompose
their amplitudes ${\cal A} $ according to the unitarity of the  CKM matrix  as
\bea{\cal A} &=& \lambda_u^s({\cal A}^{tree}_u+{\cal A}^{QCD}_u+{\cal A}^{EW}_u)
+\lambda_c^s({\cal A}^{cc}_c+{\cal A}^{non-cc}_c)\label{ampdeco}
\eea
Here  $\lambda_p^s=V_{pb}V^*_{ps}$ with $p=u,c$ and ${\cal A}^{tree}_u,{\cal A}^{QCD}_u,{\cal A}^{EW}_u$
refer to tree , $QCD$  penguin and Electroweak penguins amplitudes respectively. ${\cal A}^{cc}_c$
refers to long distance charming penguin and ${\cal A}^{non-cc}_c$ refers to contributions
from other $QCD$ and Electroweak penguins. It should be noted here that, the different amplitudes
in eq.(\ref{ampdeco}) can have zero or non zero values depending on the final state mesons.
In the SM we see that  ${\cal A}^{tree}_u\gg {\cal A}^{QCD}_u,{\cal A}^{EW}_u,{\cal A}^{non-cc}_c$
due to the hierarchy of the Wilson coefficients $C_{1,2}\gg C_{3-10}$.
 One should note that  the amplitudes ${\cal A}^{QCD}_u,{\cal A}^{EW}_u,{\cal A}^{non-cc}_c$ can receive contributions
from QCD corrections that are proportional to the large Wilson coefficients  $C_{1,2,8g}$.

 The dominant NLO QCD corrections to Wilson Coefficients
 given in Refs.\cite{Beneke:1999br,Beneke:2000ry,Chay:2003ju,Beneke:2005vv,Beneke:2006mk,Jain:2007dy}) are
taken into account in our analysis. These corrections are
important since they contribute to the strong phase required for
CP violation. In fact contributions to the strong phase from NLO
QCD corrections to ${\cal A}^{tree}_u,{\cal A}^{QCD}_u,{\cal
A}^{EW}_u$  will be suppressed roughly speaking by  a factor
$\alpha_s/\pi \times \frac{|\lambda_u^s|}{|\lambda_c^s|}\sim
0.0008$ in comparison with the strong phase of the charm penguin.
On the other hand NLO QCD corrections to $ {\cal A}^{non-cc}_c $
will be suppressed roughly speaking  by a factor $\alpha_s/\pi
\sim 0.04$ in comparison with the strong phase of the charm
penguin. Thus, in SCET, the strong phase of the charm penguin is
the dominant in all cases.

% However these
%contributions will be suppressed roughly speaking  by  a factor $\alpha_s/\pi\sim 0.04$.

 Now we consider
two cases, first case we have ${\cal A}^{tree}_u= 0$ while the second ${\cal A}^{tree}_u\neq 0$.
In the first case we can write to a good approximation, after using $\frac{|\lambda_u^s|}{|\lambda_c^s|}\sim 0.02$,
\bea{\cal A} &=& \lambda_c^s({\cal A}^{cc}_c+{\cal A}^{non-cc}_c)\label{ampdecoo}\eea
which shows that the long distance charm penguin gives the dominant contribution to the amplitude
as ${\cal A}^{non-cc}_c$ are highly suppressed by the Wilson coefficients $C_{3-10}$.
As an example for this case, the decay modes
$B^+\to\pi^+ \bar{K}^{(*)\,0}$ and $B^+\to \rho^+ \bar{K}^{0}$ where ${\cal A}^{tree}_u= 0$
and thus we expect that $Br (B^+\to\pi^+ \bar{K}^{(*)\,0})\sim Br (B^+\to \rho^+ \bar{K}^{0})$
which is clear from Table \ref{branch}.

Turning now to the second case where  ${\cal A}^{tree}_u\neq 0$,
to a good approximation we can write
\bea{\cal A} &=& \lambda_u^s{\cal A}^{tree}_u+\lambda_c^s({\cal A}^{cc}_c+{\cal A}^{non-cc}_c)\label{ampdecoo}\eea
which shows also that the long distance charm penguin gives the dominant contribution
to the amplitude, as ${\cal A}^{tree}_u $ will be suppressed by a factor $|\lambda_u^s|\sim 0.02|\lambda_c^s|$
in comparison to ${\cal A}^{cc}_c$. Thus in all cases the
long distance charm penguin gives the dominant contribution and as a consequence
the amplitude in each  decay mode will be of the same order of the long distance charming
penguin amplitude.

For decay modes which do not receive contribution from charm penguin one expects very small
branching ratios. Hence non-perturbative charming penguin plays crucial  rule in the branching ratios using SCET.

 The branching ratios of the decay modes $B\to
\pi K^*$ and $B\to \rho K$ are given in Tables \ref{branch} where the
first uncertainty in the predictions is due to the  uncertainties in SCET parameters
while the second uncertainty  is due to  the  uncertainties in the CKM
matrix elements. As can be seen from that Table, within SM, the branching ratios are in
agreements with their corresponding experimental values in most of
the decay modes.

 Turning now to the SM predictions for the CP asymmetries which are
presented in Table \ref{asym} where, as before, where the
first uncertainty in the predictions is due to the  uncertainties in SCET parameters
while the second uncertainty  is due to  the  uncertainties in the CKM
matrix elements . Clearly from the Table, the SM
predictions for the CP asymmetries of $B^+\to \pi^0 K^{*\,+}$ has different sign
in comparison with the experimental measurement and the predicted CP asymmetries
in  many of the decay modes are in agreement with the experimental measurements due to the
large errors in these measurements. Moreover, we see from the Table
that, the predicted CP asymmetry of
$\bar{B}\to \pi^0 \bar{K}^{*\,0}$ and $B^+\to\rho^0 K^+$
disagree with the experimental results within $1\sigma.$ error of the experimental data.
This can be attributed to the lack of the weak CP
violating phases as SM Wilson coefficients are real and the only
source of the weak phase is the phase of the CKM matrix.

 Note, SCET provides large strong  phases  and thus with new sources of weak
CP violation one would expect enhancement in these asymmetries.
 In the next section we consider the case of SUSY models with non
universal A terms where new sources of weak CP phases exist.

\begin{table}
\begin{center}
\begin{tabular}{|c|c|c|c|c|c|}
  \hline
    Decay channel& Exp. &  SM prediction  \\
  \hline
 $\pi^0 K^{*\,+}$  & $0.04 \pm 0.29$ & $ - 0.08_{-0.03-0.002}^{+0.03+0.002}$ \\
 \hline
    $\pi^- K^{*\,+}$ & $-0.18 \pm 0.07$ &$ -0.12_{-0.03-0.001}^{+0.04+0.01}$  \\\hline
  $ \pi^0 \bar{K}^{*\,0}$ & $-0.15\pm 0.12 $& $ - 0.01_{-0.003-0.003}^{+0.002+0.0003}$  \\\hline
  $ \pi^+ \bar{K}^{*\,0}$ &$-0.038 \pm 0.042$ & $ -0.004_{-0.001-0.0003}^{+0.001+0.001}$ \\\hline
$\rho^0 K^{+}$&$0.37\pm 0.11$ & $ 0.06_{-0.08-0.002}^{+0.07+0.002}$  \\\hline
$ \rho^+ \bar{K}^{0}$& $-0.12 \pm 0.17$&  $ - 0.005_{-0.001-0.0001}^{+0.001+0.0004}$\\\hline
$ \rho^0 \bar{K}^{0}$&$-0.02 \pm 0.27 \pm 0.08 \pm 0.06$ &$ - 0.02_{-0.01-0.001}^{+0.01+0.002}$ \\\hline
$\rho^- K^{+} $& $0.15 \pm 0.06$&  $ 0.14_{-0.11-0.01}^{+0.11+0.004}$   \\
\hline
\end{tabular}
 \end{center}
\caption{ Direct CP asymmetries of $B\to \pi K^*$ and $B\to \rho
K$ decays. As before, we list the experimental results given in
Ref.\cite{TheHeavyFlavorAveragingGroup:2010qj}. The
first uncertainty in the predictions is due to the uncertainties in SCET parameters
while the second uncertainty  is due to  the  uncertainties in the CKM
matrix elements. }\label{asym}
\end{table}

\section{SUSY contributions to the CP asymmetries of$ B\to \rho K$ and $ B\to \pi K^*$}\label{sec:SUSY}

 In this section we analyze the SUSY contributions to the CP asymmetries of
   $B^-\to \pi^-\bar{K}^{(*)\,0},B^-\to
\rho^-\bar{K}^0,\bar{B}^0\to \rho^+K^-$ and $B^-\to \rho^0 K^-$ as
their SM prediction is very small and can not accommodate  the
experimental results. In SUSY, Flavor Changing Neutral
Current(FCNC) and CP quantities are sensitive to
particular entries in the mass matrices of the scalar fermions.
Thus it is  useful to adopt a model independent-
parametrization, the so-called Mass Insertion Approximation (MIA)
where all the couplings of fermions and sfermions to neutral
gauginos are flavor diagonal~\cite{Hall:1985dx}. Denoting by $\Delta$ the
off-diagonal terms in the $(M^2_{\tilde{f}})_{AB}$ where
$\tilde{f}$ denotes any scalar fermion
and $A,B$ indicate chirality, $%
A,B=(L,R)$:

\be (M^2_{\tilde{f}})_{AB} = \left(
\begin{array}{ccc}
  (m^2_{f1})_{AB}    & (\Delta^f_{AB})_{12} & (\Delta^f_{AB})_{13}
\vspace{0.2cm} \\
(\Delta^f_{AB})_{21} &    (m^2_{f2})_{AB}   & (\Delta^f_{AB})_{23}
\vspace{0.2cm} \\
(\Delta^f_{AB})_{31} &(\Delta^f_{AB})_{32}&   (m^2_{f3})_{AB}
\end{array}\right),
\ee
 $\Delta^{IJ}_{LL} = \Delta^{JI\star}_{LL}$ and $\Delta^{IJ}_{RR}
= \Delta^{JI\star}_{RR}$, but no such relation holds for
$\Delta_{LR}$. It is often to set $ (m^2_{f1})_{AB}=
(m^2_{f2})_{AB}= (m^2_{f3})_{AB}=\tilde{m}^2$ where $\tilde{m}$ is
the average sfermion mass. The Flavour Changing structure of the
$A-B$ sfermion propagator   is exhibited by its non-diagonality
and it can be expanded as
\begin{equation}
\langle \tilde{f}_{A}^{a}\tilde{f}_{B}^{b\ast }\rangle =i(k^{2}I-\tilde{m}%
^{2}I-\Delta _{AB}^{f})_{ab}^{-1}\simeq \frac{i\delta _{ab}}{k^{2}-\tilde{m}%
^{2}}+\frac{i(\Delta
_{AB}^{f})_{ab}}{(k^{2}-\tilde{m}^{2})^{2}}+O(\Delta ^{2}),
\end{equation}%
where $a,b=(1,2,3)$ are flavor indices and $I$ is the unit matrix.
It is convenient to define a dimensionless quantity $(\delta
_{AB}^{f})_{ab}\equiv (\Delta _{AB}^{f})_{ab}/\tilde{m}^{2}.$ As long as $%
(\Delta _{AB}^{f})_{ab}$ is smaller than $\tilde{m}^{2}$ we can
consider only the first order term in $(\delta _{AB}^{f})_{ab}$ of
the sfermion propagator expansion.

 The parameters $(\delta _{AB}^{f})_{ab}$ can be constrained
through vacuum stability argument~\cite{Casas:1996de},
experimental measurements concerning FCNC and CP violating
phenomena~\cite{Gabbiani:1996hi}. Recent studies about other possible constraints
can be found in Refs.(~\cite{Crivellin:2008mq,Crivellin:2009ar,Crivellin:2010gw}).

At next leading order in $\alpha_s$ expansion, the dominant
SUSY contributions to our  decay modes are originated from diagrams mediated by the
exchange of  gluino and chargino. The complete expressions for
the gluino and chargino contributions to the Wilson coefficients can be found in Refs.
\cite{Bertolini:1990if,Gabrielli:1994ff,Gabbiani:1996hi,Buras:2000qz}.

After including SUSY contributions to the mentioned decays and keeping
the dominant terms we find

\bea A(B^-\to \pi^-\bar{K}^{(*)\,0})\times 10^7 &\simeq&- 0.0178 (\delta^d_{LL})_{23}- 6.6914
(\delta^d_{LR})_{23}- 1.5857
(\delta^d_{RL})_{23}-(0.0052 +0.0003i) (\delta^u_{LR})_{32}\nonumber\\&-&(0.0046 - 0.0003 i )
(\delta^u_{RL})_{32}+(0.3319 - 0.0612 i),\nonumber\\
 A(B^-\to \pi^0\bar{K}^{(*)\,-})\times 10^7 &\simeq& 0.0125(\delta^d_{LL})_{23}+4.7315
(\delta^d_{LR})_{23}+ 1.1212 (\delta^d_{RL})_{23}+(0.0056 -
0.0001 i) (\delta^u_{LR})_{32}\nonumber\\&-& (0.0223 - 0.0001 i )
(\delta^u_{RL})_{32}+ (0.2508 - 0.1259 i),
\nonumber\\
 A(B^0\to \pi^0\bar{K}^{(*)\,0})\times 10^7 &\simeq& -0.0127(\delta^d_{LL})_{23}-4.7315
(\delta^d_{LR})_{23}- 1.1212 (\delta^d_{RL})_{23}+(0.0094 +
0.0001 i) (\delta^u_{LR})_{32}\nonumber\\&-& (0. 0185+ 0.0001 i )
(\delta^u_{RL})_{32}+  (0.2949 - 0.0707 i),\nonumber\\
 A(B^0\to \pi^+\bar{K}^{(*)\,-})\times 10^7&\simeq& 0.0178(\delta^d_{LL})_{23}+6.6914
 (\delta^d_{LR})_{23}+ 1.5857 (\delta^d_{RL})_{23}-(0.0106+
0.0005 i) (\delta^u_{LR})_{32}\nonumber\\&-& (0.0099 - 0.0005 i )
(\delta^u_{RL})_{32}+  (0.2695 - 0.1392 i),\nonumber\\
 A(\bar{B}^-\to \rho^- K^0)\times
10^7&\simeq& 0.0043(\delta^d_{LL})_{23}+1.6190 (\delta^d_{LR})_{23}-
1.0851(\delta^d_{RL})_{23}-(0.0001 +
0.0005 i) (\delta^u_{LR})_{32}\nonumber\\&-& (0.0021 - 0.0005i )
(\delta^u_{RL})_{32}-  (0.3473 + 0.0111 i),\nonumber\\
 A(B^-\to \rho^0 K^-)\times 10^7&\simeq& -0.0031(\delta^d_{LL})_{23}-1.1448 (\delta^d_{LR})_{23}
 + 0.7673(\delta^d_{RL})_{23}-(0.0037 +0.0006 i) (\delta^u_{LR})_{32}
\nonumber\\&-&(0.0120 -0.0006 i) (\delta^u_{RL})_{32}- (0.2232 +0.0501 i),
\nonumber\\
A(\bar{B}^0\to \rho^0K^0)\times
10^7&\simeq& 0.0030(\delta^d_{LL})_{23}+1.1448 (\delta^d_{LR})_{23}-
0.7673(\delta^d_{RL})_{23}-(0.0032 + 0.0003 i) (\delta^u_{LR})_{32}\nonumber\\&-& (0.0108. - 0.0003 i )
(\delta^u_{RL})_{32} - (0.3470 + 0.0307 i),\nonumber\\
 A(B^-\to \rho^+ K^-)\times 10^7&\simeq& -0.0043(\delta^d_{LL})_{23}-1.6190 (\delta^d_{LR})_{23}
 + 1.0851(\delta^d_{RL})_{23}\nonumber\\&-&(0.0008 +0.0010 i) (\delta^u_{LR})_{32}
 -(0.0037 -0.0010 i) (\delta^u_{RL})_{32}\nonumber\\&-& (0.1723 + 0.0386 i),\label{amplitudes}\,\eea

The mass insertions $(\delta^u_{RL})_{32}$ and $(\delta^u_{LR})_{32}$ are not constrained by $b \to s \gamma$
 and so we can set them as $(\delta^u_{RL})_{32}=(\delta^u_{LR})_{32}= e^{i\delta_u}$
 where $\delta_u$ is the phase that can vary from $-\pi$ to $\pi$.  It should be noted that in order
 to have a well defined Mass Insertion Approximation scheme, it is necessary to have

\begin{figure}[tbhp]
\includegraphics[width=6.5cm,height=7cm]{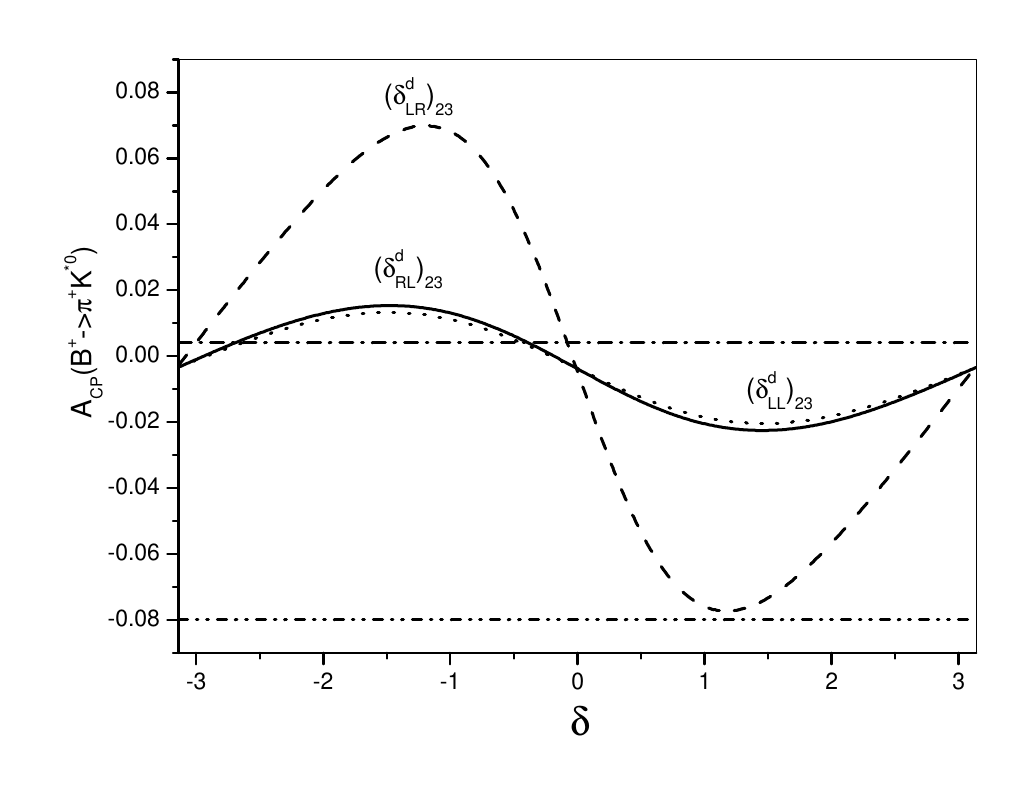}
\hspace{1.cm}
\includegraphics*[width=6.5cm,height=7cm]{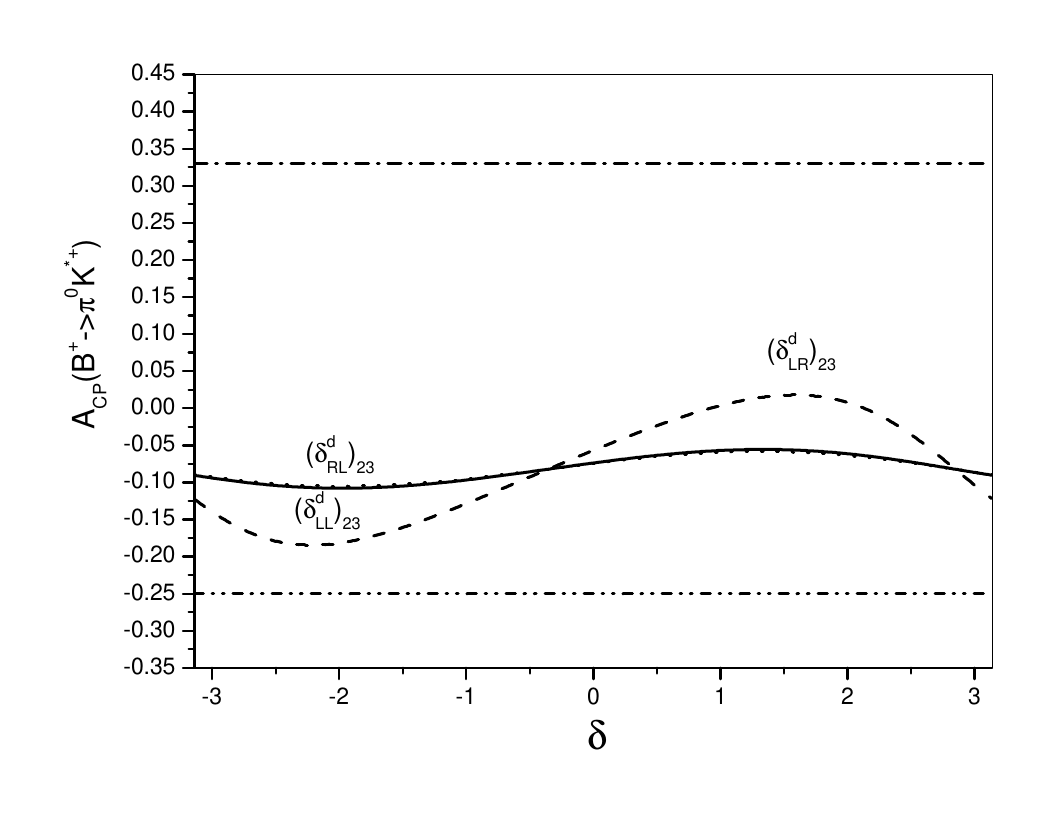}
\medskip
\caption{ CP asymmetries  versus the phase of the
$(\delta^{d}_{AB})_{23}$ where A and B denote the chirality i.e.
L, R. for 3 different mass insertions.
 The left diagram corresponds to  $A_{CP}(B^+\to \pi^+\bar{K}^{*\,0})$ while the right diagram corresponds to
 $A_{CP}(B^+\to \pi^0 K^{*\,+})$. In both diagrams we take only one mass insertion
per time and vary the phase of  from $-\pi$ to $\pi$. The horizontal lines
in both diagrams represent the experimental measurement to $1\sigma.$} \label{singlemas}
\end{figure}

 $|(\delta^f_{AB})_{ab} | < 1$
 but here in order to maximize the SUSY CP-violating contributions we take it of order one.
 Applying $b \to s \gamma$ constraints
 leads to the following parametrization~\cite{Huitu:2009st}

 \begin{equation}
 (\delta^d_{LL})_{23}= e^{i\delta_d}~~~~~~~~~~~ (\delta^d_{LR})_{23}=(\delta^d_{RL})_{23}= 0.01 e^{i\delta}\label{masinscons}
 \end{equation}

In the following we present our results for the CP asymmetries. In
our analysis we consider two scenarios, the first one with a
single mass insertion where we keep only one mass insertion per time and
take the other mass insertions to be zero and the second scenario with
two mass insertions will be considered only in the cases when
one single mass insertion is not sufficient to accommodate the experimental measurement.
After setting the different  mass insertions as mentioned above, we see from Eq.(\ref{amplitudes})
that, the terms that contain the mass insertions $(\delta^{u}_{RL})_{32}$ and $(\delta^{u}_{LR})_{32}$
will be small in comparison  with the other terms and thus we expect that their contributions
to the asymmetries will be small. These terms are obtained from diagrams mediated by the chargino
exchange and thus we see that gluino contributions give the dominant contributions as known in the literature.

 We start our analysis of the direct CP asymmetries by considering the first scenario in which
 we take only one mass insertion corresponding to the gluino mediation
 and set the others to be zero.

 \begin{figure}[tbhp]
\includegraphics[width=6.5cm,height=7cm]{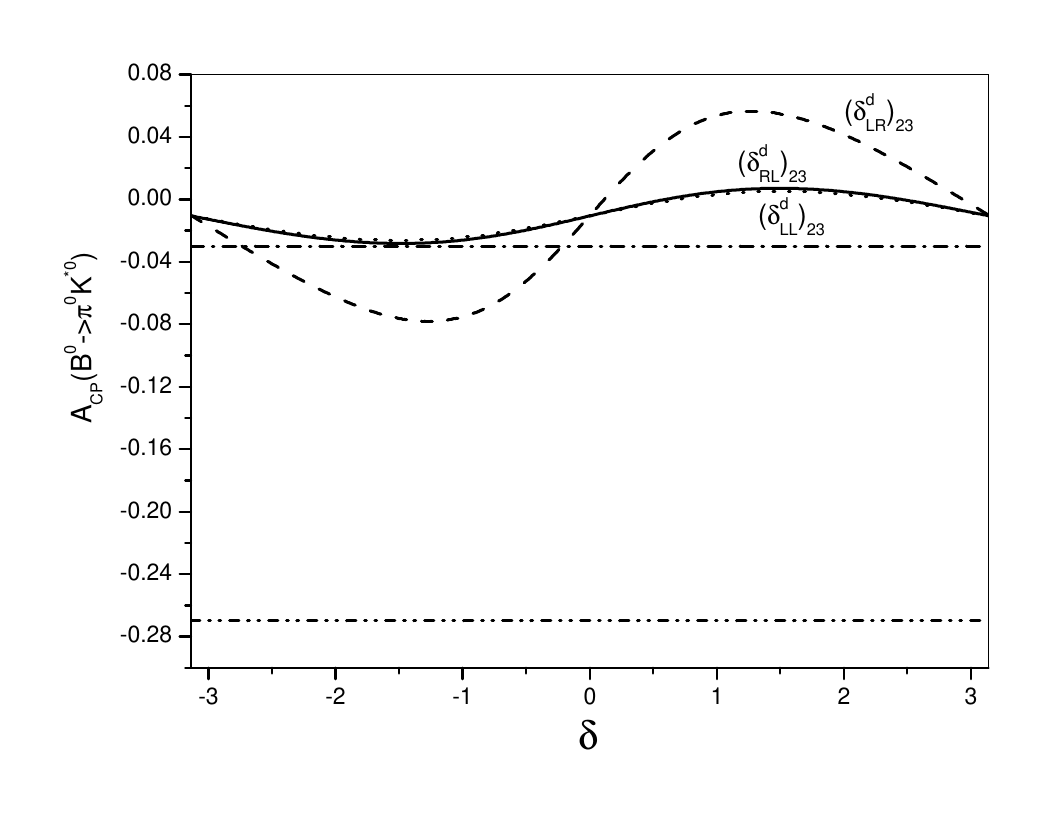}
\hspace{1.cm}
\includegraphics*[width=6.5cm,height=7cm]{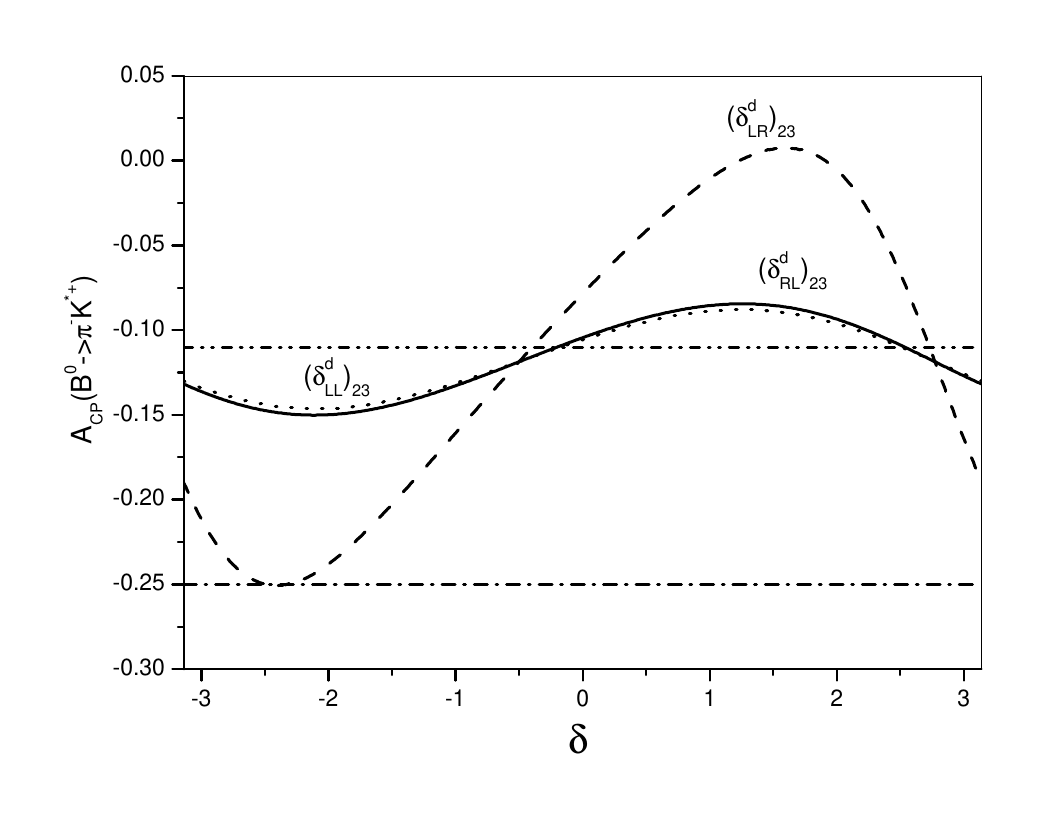}
\medskip
\caption{CP asymmetries  versus the
phase of the $(\delta^{d}_{AB})_{23}$ where A and B denote the chirality i.e. L, R. for 3 different mass insertions.
 The left diagram corresponds to  $A_{CP}(B^0\to \pi^0\bar{K}^{*\,0})$ while the right diagram corresponds to
$A_{CP}(B^0\to \pi^-K^{*\,+})$. In both diagrams we take only one mass insertion
per time and vary the phase of  from $-\pi$ to $\pi$. The horizontal lines
in both diagrams represent the experimental measurement to $1\sigma$.} \label{singlemas2}
\end{figure}

\begin{figure}[tbhp]
\includegraphics[width=6.5cm,height=7cm]{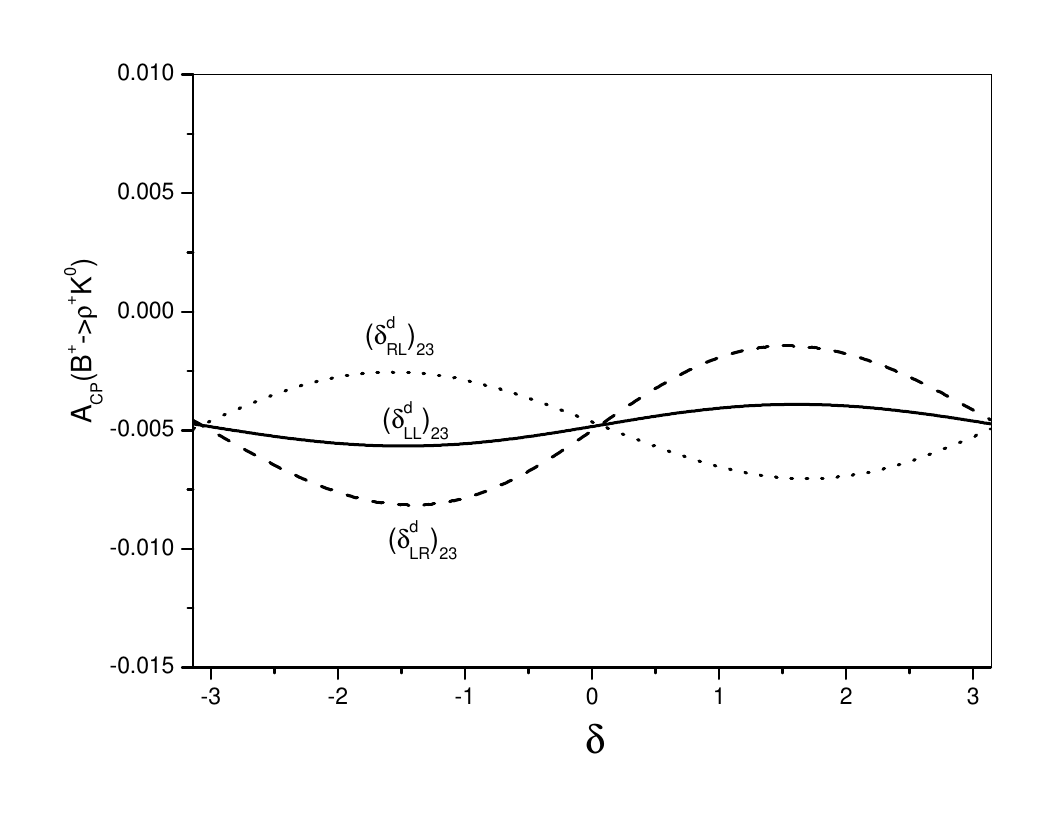}
\hspace{1.cm}
\includegraphics*[width=6.5cm,height=7cm]{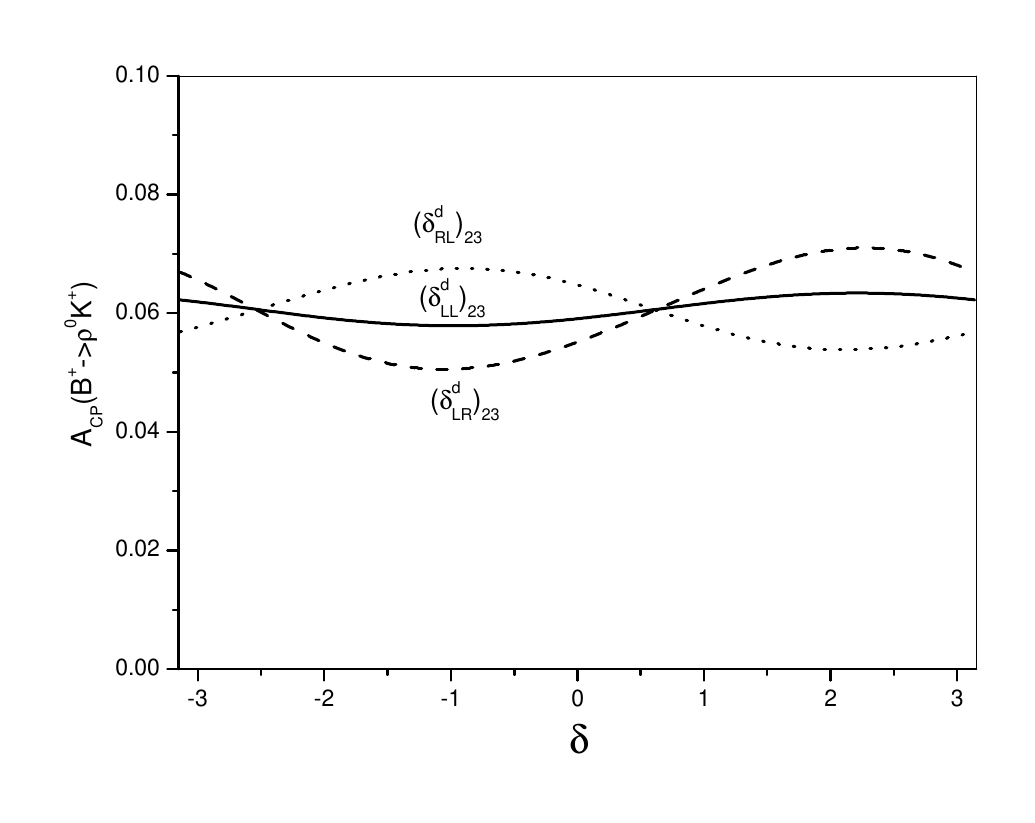}
\medskip
\caption{CP asymmetries  versus the phase of the
$(\delta^{d}_{AB})_{23}$ where A and B denote the chirality i.e.
L, R. for 3 different mass insertions.
 The left diagram corresponds to  $A_{CP}(B^+\to \rho^+ K^{0})$ while the right diagram corresponds to
$A_{CP}(B^+\to \rho^0 K^{+})$. In both diagrams we take only one
mass insertion per time and vary the phase of  from $-\pi$ to
$\pi$.} \label{singlemas3}
\end{figure}

 After substituting the mass insertions given in eq.(\ref{masinscons}) in  eq.(\ref{amplitudes})
we find that the first and third terms in the amplitudes $B^+\to \pi^+\bar{K}^{*\,0}$
and $B^+\to \pi^0 K^{*\,+}$ will be approximately equal and both of them will be smaller than the second term.
 As a consequence, one predicts that the asymmetries generated by the mass
insertions $(\delta^d_{LL})_{23}$ and $(\delta^d_{RL})_{23}$ will be equal and in the same time
these asymmetries will be smaller than the case of using $(\delta^d_{LR})_{23}$ which can be
seen from Fig.\ref{singlemas}. In that Figure,we plot the  CP asymmetries,  $A_{CP}(B^+\to \pi^+\bar{K}^{*\,0})$ and
 $A_{CP}(B^+\to \pi^0 K^{*\,+})$  versus the phase of the $(\delta^{d}_{AB})_{32}$
 where A and B denote the chirality i.e. L and R. for 3 different mass insertions. The horizontal lines
in both diagrams represent the experimental measurements to $1\sigma.$
  As can be seen from  Figure\ref{singlemas}~left, for all gluino mass insertions, the value of
the CP asymmetry $A_{CP}(B^+\to \pi^+\bar{K}^{*\,0})$ is enhanced  to accommodate  the
experimental measurement of the asymmetry  within $1\sigma$ for many values
of the phase of the mass insertions. On the other hand,Figure \ref{singlemas}right shows that
 the CP asymmetry  $A_{CP}(B^+\to \pi^0 K^{*\,+})$ is enhanced  to accommodate  the
experimental measurement within $1\sigma$ for all values of the phase of the mass insertions. The point
we stress here is that SUSY Wilson coefficients provide source of large weak phases, which are needed for
accommodation of CP asymmetries.

   In Fig.\ref{singlemas2}  we plot the two asymmetries, $A_{CP}(B^0\to \pi^0\bar{K}^{*\,0})$ and
 $A_{CP}(B^0\to \pi^-K^{*\,+})$ versus the phase of the $(\delta^{d}_{AB})_{32}$  as before.
 As can be seen from  Fig.\ref{singlemas2} left, $A_{CP}(B^0\to \pi^0\bar{K}^{*\,0})$  lies within
$1\sigma$ range of its experimental value for  many values of the phase of the mass
insertion $(\delta^{d}_{LR})_{23}$ only. The reason for that is as before,(see eq.(\ref{amplitudes})) the two mass insertions
$(\delta^d_{LL})_{23}$ and $(\delta^d_{RL})_{23}$ will give equal contributions to the CP asymmetries which
 will be smaller than the case of using $(\delta^d_{LR})_{23}$.
 On the other hand, Fig.\ref{singlemas2} right, we see that  $A_{CP}(B^0\to \pi^-K^{*\,+})$ can
be accommodated within $1\sigma$ for many values of the phase of the three gluino mass insertions.

\begin{figure}[tbhp]
\includegraphics[width=6.5cm,height=7cm]{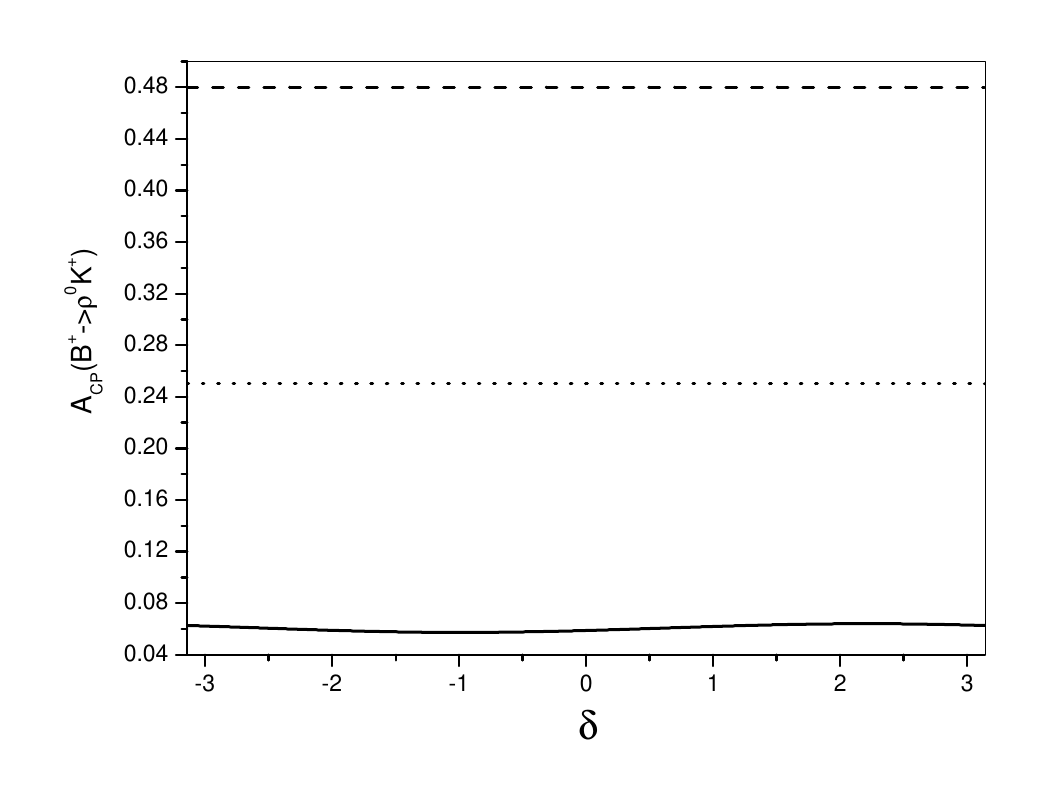}
\hspace{1.cm}
\includegraphics*[width=6.5cm,height=7cm]{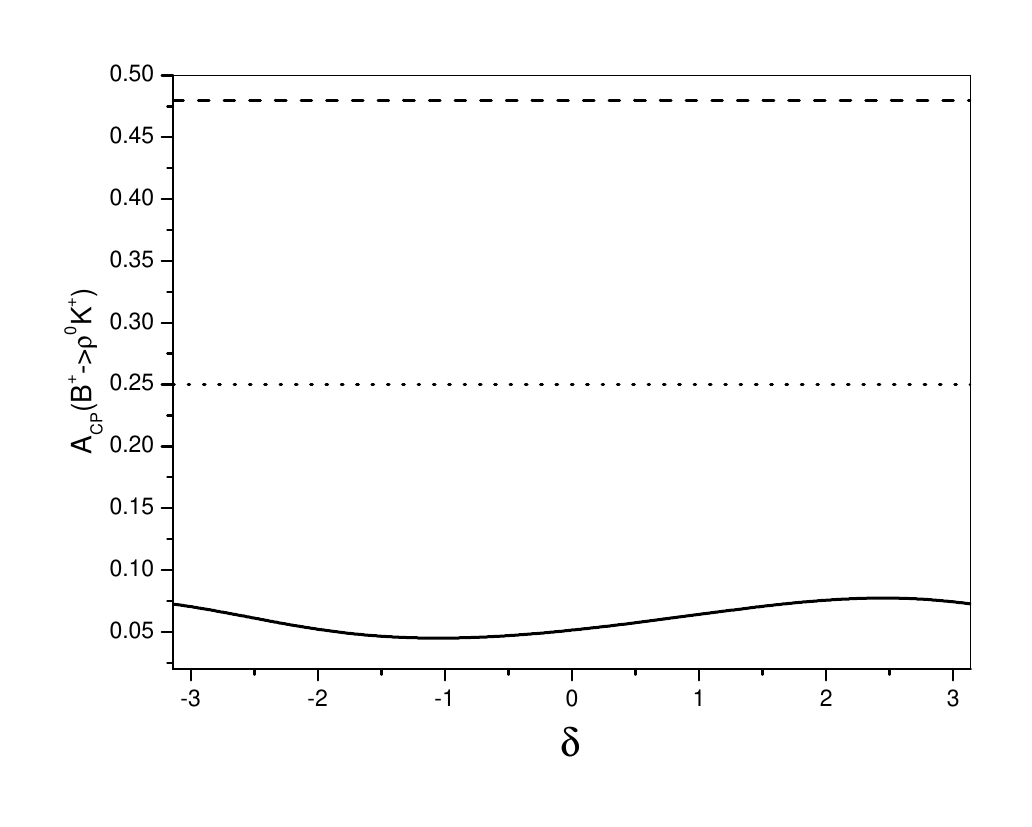}
\medskip
\caption{CP asymmetry of $A_{CP}(B^+\to \rho^0 K^{+})$ versus the
phase of the mass insertion  for 2 different mass insertions.
 The left diagram correspond to gluino contributions where we keep
 the two mass insertions $(\delta^{d}_{LR})_{23}$ and $(\delta^{d}_{RL})_{23}$
 and set the other mass insertions to zero.   The right diagram correspond to
both gluino and chargino contributions where we keep
 the two mass insertions $(\delta^{d}_{LR})_{23}$ and $(\delta^{u}_{RL})_{32}$
 and set the other mass insertions to zero. In both diagrams we assume that the two mass insertion
have equal phases and we vary the phase  from $-\pi$ to $\pi$. The
horizontal lines in both diagrams represent the experimental
measurements to $1\sigma$.} \label{singlemas4}
\end{figure}

Finally we discuss the CP asymmetries of the decay modes $B^+\to \rho^+ K^{0}$ and $B^+\to \rho^0 K^{+}$.
  After substituting the mass insertions given in eq.(\ref{masinscons}) in  eq.(\ref{amplitudes}),
we find that the first and third terms in the  amplitudes $B^+\to \rho^+ K^{0}$ and $B^+\to \rho^0 K^{+}$
will be no longer equal as previous cases and thus we expect their contributions to the asymmetries will  be different
which can be seen from  Fig.(\ref{singlemas3}) where, as before, we plot
$A_{CP}(B^+\to \rho^+ K^{0})$ and  $A_{CP}(B^+\to \rho^0 K^{+})$
versus the phase of the $(\delta^{d}_{AB})_{23}$. In Fig.(\ref{singlemas3}) we do not show the
horizontal lines representing the $1\sigma$ range of the experimental measurement as the three curves
of the $A_{CP}(B^+\to \rho^+ K^{0})$ corresponding to the three gluino mass insertions totally
lie in this $1\sigma$ range for all values of the phase of the mass insertions.
 On the other hand, Fig.\ref{singlemas3} right, we see that
$A_{CP}(B^+\to \rho^0 K^{+})$  can not be accommodated within $1\sigma$ for any value of the phase of
all gluino mass insertions. This motivates us to consider the second scenario with two mass insertions.

\begin{figure}[tbhp]
\includegraphics[width=6.5cm,height=7cm]{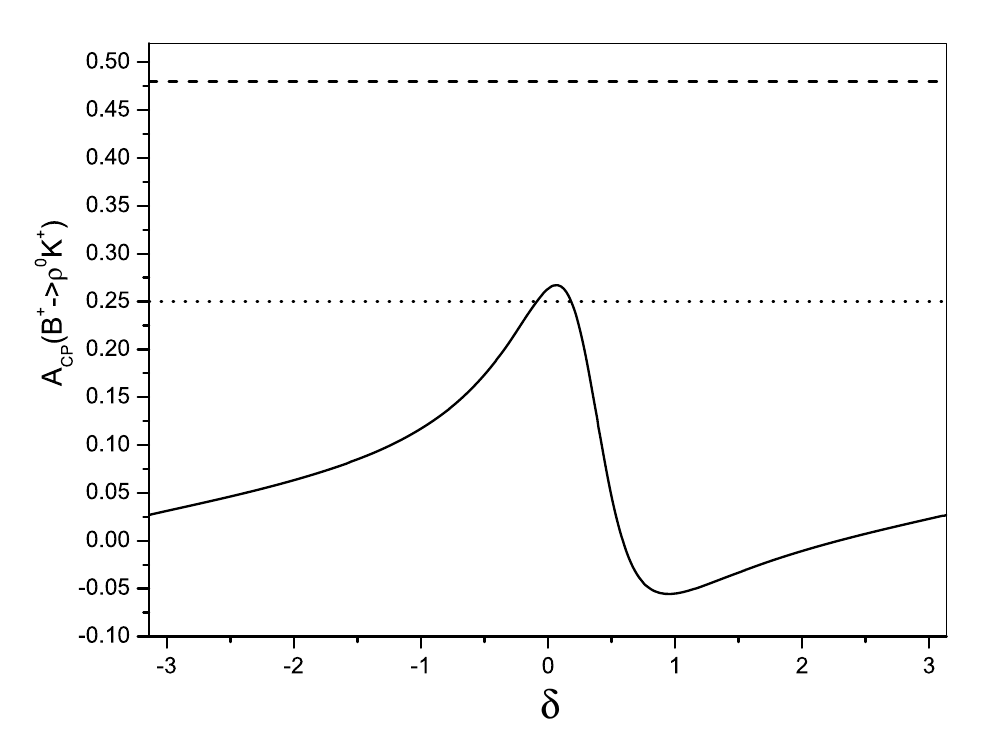}
\medskip
\caption{CP asymmetry of $A_{CP}(B^+\to \rho^0 K^{+})$ versus the
phase of the mass insertion  for 2 different mass insertions
correspond to gluino contributions where we keep the two mass
insertions $(\delta^{d}_{LR})_{23}$ and $(\delta^{d}_{LL})_{23}$
and set the other mass insertions to zero.  We assume that the two
mass insertion have equal phases and we vary the phase  from
$-\pi$ to $\pi$. The horizontal lines in the diagram represent the
experimental measurements to $1\sigma$.} \label{singlemas5}
\end{figure}

In Fig.\ref{singlemas4}, we plot the CP asymmetry, $A_{CP}(B^+\to \rho^0 K^{+})$ versus the
phase of the mass insertion  for 2 different mass insertions.
The left diagram correspond to gluino contributions where we keep
the two mass insertions $(\delta^{d}_{LR})_{23}$ and $(\delta^{d}_{RL})_{23}$
 and set the other mass insertions to zero.   The right diagram correspond to
both gluino and chargino contributions where we keep  the two mass insertions $(\delta^{d}_{LR})_{23}$ and $(\delta^{u}_{RL})_{32}$
 and set the other mass insertions to zero. In both diagrams we assume that the two mass insertion
have equal phases and we vary the phase  from $-\pi$ to $\pi$. As
before, the horizontal lines in both diagrams represent the
experimental measurement to $1\sigma$. As can be seen from
Fig.\ref{singlemas4} left, two gluino mass insertions can not
accommodate the experimental measurement for any value of the
phase of the mass insertion. On the other hand from
Fig.\ref{singlemas4} right, two mass insertions one corresponding
to chargino contribution and the other corresponding to  gluino
contribution can not  accommodate the experimental measurements.
We find that in order to accommodate the CP symmetry in this case
the Wilson coefficient $C^{\tilde{g}}_{9}$ should be increased at
least by a factor $-6\pi/\alpha$ without violating any constraints
on the SUSY parameter space. We show the corresponding diagram in
Fig.\ref{singlemas5}.

\section{Conclusion}\label{sec:conclusion}

Within Soft Collinear Effective Theory, we extend the Standard
Model analysis of the $B \to \pi K^*$ and $B\to\rho K$ asymmetries
to include the next leading order QCD corrections. We find that,
even with QCD correction, the Standard Model predictions can not
accommodate the direct CP asymmetries in these decay modes.

We have analyzed the  SUSY contributions to the direct CP
asymmetries of  the decay modes $ B\to \rho K$ and  $ B\to \pi
K^*$ using the Mass Insertion Approximation. Contrarily to SM, our
results show that  these direct CP asymmetries can be
significantly enhanced by the SUSY contributions mediated by
gluino exchange and thus accommodate the experimental results.

\section*{Acknowledgement}

Gaber Faisel's  work is supported by the National Science Council of
R.O.C. under grants NSC 99-2112-M-008-003-MY3 and NSC
99-2811-M-008-085.D.D. has been supported by PROMEP and DINPO project from Guanajuato University.
%\newpage

%%%%%%%%%%%%%%%%%%%%%%%%%%%%%%%%%%%%%%%%%%%%%%%%%%%%%%%%%%


\begin{thebibliography}{99}




\bibitem{TheHeavyFlavorAveragingGroup:2010qj}
  The Heavy Flavor Averaging Group {\it et al.},
  %``Averages of b-hadron, c-hadron, and tau-lepton Properties,''
  arXiv:1010.1589 [hep-ex].
  %%CITATION = ARXIV:1010.1589;%%

\bibitem{Beneke:2003zv}
  M.~Beneke and M.~Neubert,
  %``QCD factorization for B --> P P and B --> P V decays,''
  Nucl.\ Phys.\  B {\bf 675}, 333 (2003)
  [arXiv:hep-ph/0308039].
  %%CITATION = NUPHA,B675,333;%%



\bibitem{Lu:2000hj}
  C.~D.~Lu and M.~Z.~Yang,
  %``B --> pi rho, pi omega decays in perturbative QCD approach,''
  Eur.\ Phys.\ J.\  C {\bf 23}, 275 (2002)
  [arXiv:hep-ph/0011238].
  %%CITATION = EPHJA,C23,275;%%


%\cite{Liu:2005mm}
\bibitem{Liu:2005mm}
  X.~Liu, H.~s.~Wang, Z.~j.~Xiao, L.~Guo and C.~D.~Lu,
  %``Branching ratio and CP asymmetry of B --> rho eta(') decays in the
  %perturbative QCD approach,''
  Phys.\ Rev.\  D {\bf 73}, 074002 (2006)
  [arXiv:hep-ph/0509362].
  %%CITATION = PHRVA,D73,074002;%%


%\cite{Guo:2007vw}
\bibitem{Guo:2007vw}
  D.~Q.~Guo, X.~F.~Chen and Z.~J.~Xiao,
  %``B0 --> omega eta(') and Phi eta(') decays in the perturbative QCD
  %approach,''
  Phys.\ Rev.\  D {\bf 75}, 054033 (2007)
  [arXiv:hep-ph/0702110].
  %%CITATION = PHRVA,D75,054033;%%
%\cite{Guo:2006uq}
\bibitem{Guo:2006uq}
  L.~Guo, Q.~g.~Xu and Z.~j.~Xiao,
  %``B --> K K* decays in the perturbative QCD approach,''
  Phys.\ Rev.\  D {\bf 75}, 014019 (2007)
  [arXiv:hep-ph/0609005].
  %%CITATION = PHRVA,D75,014019;%%

\bibitem{Wang:2008rk}
  W.~Wang, Y.~M.~Wang, D.~S.~Yang and C.~D.~Lu,
  %``Charmless Two-body $B(B_s)\to VP$ decays In
  %Soft-Collinear-Effective-Theory,''
  Phys.\ Rev.\  D {\bf 78}, 034011 (2008)
  [arXiv:0801.3123 [hep-ph]].
  %%CITATION = PHRVA,D78,034011;%%

%\cite{Chiang:2009hd}
\bibitem{Chiang:2009hd}
  C.~W.~Chiang and D.~London,
  %``Looking for New Physics in B --> K^* \pi and B --> \rho K Decays,''
  Mod.\ Phys.\ Lett.\  A {\bf 24}, 1983 (2009)
  [arXiv:0904.2235 [hep-ph]].
  %%CITATION = MPLAE,A24,1983;%%



%\cite{Delepine:2008zzb}
\bibitem{Delepine:2008zzb}
  D.~Delepine, G.~Faisel, S.~Khalil and M.~Shalaby,
  %``Supersymmetric contributions to CP asymmetry in tau-decays,''
  Int.\ J.\ Mod.\ Phys.\  A {\bf 22}, 6011 (2007).
  %%CITATION = IMPAE,A22,6011;%%


%\cite{Delepine:2007qg}
\bibitem{Delepine:2007qg}
  D.~Delepine, G.~Faisel and S.~Khalil,
  %``SUSY R parity violation and CP asymmetry in semi-leptonic tau-decays,''
  Phys.\ Rev.\  D {\bf 77}, 016003 (2008)
  [arXiv:0710.1441 [hep-ph]].
  %%CITATION = PHRVA,D77,016003;%%


%\cite{Delepine:2006fv}
\bibitem{Delepine:2006fv}
  D.~Delepine, G.~Faisl, S.~Khalil and G.~L.~Castro,
  %``Supersymmetry and CP violation in |Delta(S)| = 1 tau decays,''
  Phys.\ Rev.\  D {\bf 74}, 056004 (2006)
  [arXiv:hep-ph/0608008].
  %%CITATION = PHRVA,D74,056004;%%






\bibitem{Bauer:2000ew}
  C.~W.~Bauer, S.~Fleming and M.~E.~Luke,
  %``Summing Sudakov logarithms in B --> X/s gamma in effective field  theory,''
  Phys.\ Rev.\  D {\bf 63}, 014006 (2000)
  [arXiv:hep-ph/0005275].
  %%CITATION = PHRVA,D63,014006;%%


\bibitem{Bauer:2000yr}
  C.~W.~Bauer, S.~Fleming, D.~Pirjol and I.~W.~Stewart,
  %``An effective field theory for collinear and soft gluons: Heavy to light
  %decays,''
  Phys.\ Rev.\  D {\bf 63}, 114020 (2001)
  [arXiv:hep-ph/0011336].
  %%CITATION = PHRVA,D63,114020;%%



  \bibitem{Chay:2003zp}
  J.~Chay and C.~Kim,
  %``Factorization of B decays into two light mesons in soft-collinear
  %effective theory,''
  Phys.\ Rev.\ D {\bf 68}, 071502 (2003)
  [arXiv:hep-ph/0301055].
  %%CITATION = HEP-PH 0301055;%%

\cite{Chay:2003ju}
\bibitem{Chay:2003ju}
  J.~Chay and C.~Kim,
  %``Nonleptonic B decays into two light mesons in soft-collinear effective
  %theory,''
  Nucl.\ Phys.\ B {\bf 680}, 302 (2004)  [arXiv:hep-ph/0301262].
  %%CITATION = HEP-PH 0301262;%%






%\cite{Fleming:2009fe}
\bibitem{Fleming:2009fe}
  S.~Fleming,
  %``Soft Collinear Effective Theory: An Overview,''
  PoS E {\bf FT09}, 002 (2009)
  [arXiv:0907.3897 [hep-ph]].
  %%CITATION = POSCI,EFT09,002;%%







  \bibitem{Bauer:2002aj}
  C.~W.~Bauer, D.~Pirjol and I.~W.~Stewart,
  %``Factorization and endpoint singularities in heavy-to-light decays,''
  Phys.\ Rev.\  D {\bf 67}, 071502 (2003)
  [arXiv:hep-ph/0211069].
  %%CITATION = PHRVA,D67,071502;%%

  \bibitem{Jain:2007dy}
  A.~Jain, I.~Z.~Rothstein and I.~W.~Stewart,
  %``Penguin Loops for Nonleptonic B-Decays in the Standard Model: Is there a
  %Penguin Puzzle?,''
  arXiv:0706.3399 [hep-ph].
  %%CITATION = ARXIV:0706.3399;%%

  \bibitem{Bauer:2004tj}
  C.~W.~Bauer, D.~Pirjol, I.~Z.~Rothstein and I.~W.~Stewart,
  %``B $\to$ M(1) M(2): Factorization, charming penguins, strong phases, and
  %polarization,''
  Phys.\ Rev.\ D {\bf 70}, 054015 (2004)
  [arXiv:hep-ph/0401188].
  %%CITATION = HEP-PH 0401188;%%


\bibitem{Beneke:2000ry}
  M.~Beneke, G.~Buchalla, M.~Neubert and C.~T.~Sachrajda,
  %``QCD factorization for exclusive, non-leptonic B meson decays: General
  %arguments and the case of heavy-light final states,''
  Nucl.\ Phys.\  B {\bf 591}, 313 (2000)
  [arXiv:hep-ph/0006124].
  %%CITATION = NUPHA,B591,313;%%


\bibitem{Beneke:1999br}
  M.~Beneke, G.~Buchalla, M.~Neubert and C.~T.~Sachrajda,
  %``{QCD} factorization for B --> pi pi decays: Strong phases and CP  violation
  %in the heavy quark limit,''
  Phys.\ Rev.\ Lett.\  {\bf 83}, 1914 (1999)
  [arXiv:hep-ph/9905312].
  %%CITATION = PRLTA,83,1914;%%





\bibitem{Beneke:2005vv}
  M.~Beneke and S.~Jager,
  %``Spectator scattering at NLO in non-leptonic B decays: Tree amplitudes,''
  Nucl.\ Phys.\  B {\bf 751}, 160 (2006)
  [arXiv:hep-ph/0512351].
  %%CITATION = NUPHA,B751,160;%%

\bibitem{Beneke:2006mk}
  M.~Beneke and S.~Jager,
  %``Spectator scattering at NLO in non-leptonic B decays: Leading penguin
  %amplitudes,''
  Nucl.\ Phys.\  B {\bf 768}, 51 (2007)
  [arXiv:hep-ph/0610322].
  %%CITATION = NUPHA,B768,51;%%




\bibitem{Arnesen:2006vb}
  C.~M.~Arnesen, Z.~Ligeti, I.~Z.~Rothstein and I.~W.~Stewart,
  %``Power Corrections in Charmless Nonleptonic B-Decays: Annihilation is
  %Factorizable and Real,''
  arXiv:hep-ph/0607001.
  %%CITATION = HEP-PH/0607001;%%

  \bibitem{Hardmeier:2003ig}
  A.~Hardmeier, E.~Lunghi, D.~Pirjol and D.~Wyler,
  %``Subleading collinear operators and their matrix elements,''
  Nucl.\ Phys.\  B {\bf 682}, 150 (2004)
  [arXiv:hep-ph/0307171].
  %%CITATION = NUPHA,B682,150;%%

  \bibitem{Keum:2000ph}
  Y.~Y.~Keum, H.~n.~Li and A.~I.~Sanda,
  %``Fat penguins and imaginary penguins in perturbative QCD,''
  Phys.\ Lett.\  B {\bf 504}, 6 (2001)
  [arXiv:hep-ph/0004004].
  %%CITATION = PHLTA,B504,6;%%

\bibitem{Lu:2000em}
  C.~D.~Lu, K.~Ukai and M.~Z.~Yang,
  %``Branching ratio and CP violation of B --> pi pi decays in perturbative  QCD
  %approach,''
  Phys.\ Rev.\  D {\bf 63}, 074009 (2001)
  [arXiv:hep-ph/0004213].
  %%CITATION = PHRVA,D63,074009;%%

\bibitem{Beneke:2001ev}
  M.~Beneke, G.~Buchalla, M.~Neubert and C.~T.~Sachrajda,
  %``QCD factorization in B --> pi K, pi pi decays and extraction of
  %Wolfenstein parameters,''
  Nucl.\ Phys.\  B {\bf 606}, 245 (2001)
  [arXiv:hep-ph/0104110].
  %%CITATION = NUPHA,B606,245;%%




\bibitem{Kagan:2004uw}
  A.~L.~Kagan,
  %``Polarization in B --> V V decays,''
  Phys.\ Lett.\  B {\bf 601}, 151 (2004)
  [arXiv:hep-ph/0405134].
  %%CITATION = PHLTA,B601,151;%%



\bibitem{Manohar:2006nz}
  A.~V.~Manohar and I.~W.~Stewart,
  %``The zero-bin and mode factorization in quantum field theory,''
  Phys.\ Rev.\  D {\bf 76}, 074002 (2007)
  [arXiv:hep-ph/0605001].
  %%CITATION = PHRVA,D76,074002;%%


\bibitem{BBL}  G. Buchalla, A. J. Buras and M. E. Lautenbacher, Rev.
Mod. Phys {\bf 68}, 1230 (1996) [arXiv:hep-ph/9512380].











\bibitem{Hall:1985dx}
  L.~J.~Hall, V.~A.~Kostelecky and S.~Raby,
  %``New Flavor Violations In Supergravity Models,''
  Nucl.\ Phys.\  B {\bf 267}, 415 (1986).
  %%CITATION = NUPHA,B267,415;%%







 \bibitem{Casas:1996de}
  J.~A.~Casas and S.~Dimopoulos,
  %``Stability bounds on flavor-violating trilinear soft terms in the MSSM,''
  Phys.\ Lett.\  B {\bf 387}, 107 (1996)
  [arXiv:hep-ph/9606237].
  %%CITATION = PHLTA,B387,107;%%


\bibitem{Gabbiani:1996hi}
  F.~Gabbiani, E.~Gabrielli, A.~Masiero and L.~Silvestrini,
  %``A complete analysis of FCNC and CP constraints in general SUSY extensions
  %of the standard model,''
  Nucl.\ Phys.\  B {\bf 477}, 321 (1996)
  [arXiv:hep-ph/9604387].
  %%CITATION = NUPHA,B477,321;%%




%\cite{Crivellin:2008mq}
\bibitem{Crivellin:2008mq}
  A.~Crivellin and U.~Nierste,
  %``Supersymmetric renormalisation of the CKM matrix and new constraints on the
  %squark mass matrices,''
  Phys.\ Rev.\  D {\bf 79}, 035018 (2009)
  [arXiv:0810.1613 [hep-ph]].
  %%CITATION = PHRVA,D79,035018;%%


 %\cite{Crivellin:2009ar}
\bibitem{Crivellin:2009ar}
  A.~Crivellin and U.~Nierste,
  %``Chirally enhanced corrections to FCNC processes in the generic MSSM,''
  Phys.\ Rev.\  D {\bf 81}, 095007 (2010)
  [arXiv:0908.4404 [hep-ph]].
  %%CITATION = PHRVA,D81,095007;%%





%\cite{Crivellin:2010gw}
\bibitem{Crivellin:2010gw}
  A.~Crivellin and J.~Girrbach,
  %``Constraining the MSSM sfermion mass matrices with light fermion masses,''
  Phys.\ Rev.\  D {\bf 81}, 076001 (2010)
  [arXiv:1002.0227 [hep-ph]].
  %%CITATION = PHRVA,D81,076001;%%







%\cite{Buras:2000qz}
\bibitem{Buras:2000qz}
  A.~J.~Buras, P.~Gambino, M.~Gorbahn, S.~Jager and L.~Silvestrini,
  %``epsilon'/epsilon and Rare K and B Decays in the MSSM,''
  Nucl.\ Phys.\  B {\bf 592}, 55 (2001)
  [arXiv:hep-ph/0007313].
  %%CITATION = NUPHA,B592,55;%%









  %\cite{Bertolini:1990if}
\bibitem{Bertolini:1990if}
  S.~Bertolini, F.~Borzumati, A.~Masiero and G.~Ridolfi,
  %``Effects of supergravity induced electroweak breaking on rare $B$ decays and
  %mixings,''
  Nucl.\ Phys.\  B {\bf 353}, 591 (1991).
  %%CITATION = NUPHA,B353,591;%%




%\cite{Gabrielli:1994ff}
\bibitem{Gabrielli:1994ff}
  E.~Gabrielli and G.~F.~Giudice,
  %``Supersymmetric corrections to epsilon prime / epsilon at the leading order
  %in QCD and QED,''
  Nucl.\ Phys.\  B {\bf 433}, 3 (1995)
  [Erratum-ibid.\  B {\bf 507}, 549 (1997)]
  [arXiv:hep-lat/9407029].
  %%CITATION = NUPHA,B433,3;%%


\bibitem{Huitu:2009st}
  K.~Huitu and S.~Khalil,
  %``New Physics contribution to $B \to K \pi$ decays in SCET,''
  Phys.\ Rev.\  D {\bf 81}, 095008 (2010)
  [arXiv:0911.1868 [hep-ph]].
  %%CITATION = PHRVA,D81,095008;%%








\end{thebibliography}
\end{document}